# Superconductivity controlled Bulk magnetism

Biswajit Dutta[1], Sonam Bhakat[1], Pushpak Banerjee[1] & Avradeep Pal[1,2]*

[1] Department of Metallurgical Engineering and Materials Science, Indian Institute of Technology, Bombay, Powai, Maharashtra – 400076.

[2] Centre of Excellence in Quantum Information, Computation, Science and Technology, Indian Institute of Technology Bombay, Powai, Mumbai 400076, India

*Abstract: Ferromagnetism's ability to influence superconducting order is well known and well established[1–7], but the converse phenomena remains relatively less explored. Theoretical work on the subject includes Anderson and Suhl's prediction[8] of a crypto-ferromagnetic (CFM) state, and De Gennes's proposal of two ferromagnetic insulators exchange coupled through a superconductor[9]. In this study, we present compelling evidence of co-existence of both phenomena in a superconducting spin valve (SSV) system. We demonstrate that superconducting exchange coupling (SEC) enables reliable bistable states, and the coexistence of SEC and CFM leads to a wide range of reproducible zero field micro-magnetic states in the SSV, which are a function of the strength of the superconducting state. These micromagnetic states can in turn influence the superconducting state, leading to multiple reproducible and non-volatile resistance states; thus paving the way for a novel direction in cryogenic in-memory computing.*

The ability of a ferromagnet to alter the conventional s wave superconducting (S) order parameter has been a well-studied problem[1–7]. However, the reverse phenomena – that of superconducting state affecting the ferromagnetic (F) order parameter, although having theoretically preceded the former[8,9], had remained experimentally elusive until recently[10]. The initial prediction of Anderson and Suhl[8] of the 'crypto-ferromagnetic state' hinted at the possibility of formation of ultra-small domains in a magnetic superconductor. De Gennes predicted the possibility of a superconductor mediated interaction between two ferromagnetic insulators (FI), and this was recently demonstrated in GdN/Nb/GdN spin valves, where loosely coupled surface spins at Nb/GdN interfaces were attributed as the exchange coupled spins through the superconductor[10]. In this work, we demonstrate the first compelling evidence of superconductivity controlled bulk micromagnetic states in GdN/ Nb/ GdN Superconducting Spin Valves (SSVs). As a result, the remnant magnetisation of the SSV can be precisely controlled by maintaining various superconducting ground states along with small magnetic perturbations, and this translates to several non-volatile resistance states between completely on and off resistance states of the SSV, thereby demonstrating the potential of cryogenic in-memory computing[11] with such devices.

The colour plot in FIG. 1a depicts the variation of resistivity of SV1 spin-valve (5nm GdN/8.5nm Nb/ 3nm GdN) in response to temperature and magnetic field. When the magnetic field is lowered below the ferromagnet's saturation field, a sudden change in the superconducting transition temperature ($T_C$) is observed, which is typically a behaviour of conventional SSV[12,13]. The variation of the superconducting transition temperature ($\Delta T_C$) of 0.8K between the parallel and antiparallel state is observed in this SSV. The resistive isotherm (RH, FIG. 1b) at 2.8 K is extracted from the color plot, depicts switching fields $H_{c1} = \pm 5$ mT (5nm GdN) and $H_{c2} = \pm 8$ mT (3nm GdN). These switching fields are well matched with the coercive fields of the 3 nm and 5 nm GdN layer confirmed from the magnetic isotherm (MH), shown in FIG. 1c and such observations constitute standard Superconducting Exchange Coupling (SEC) SSV features reported previously.

A bistable switching phenomenon can be attained in these SSVs as depicted in FIG. 1c, where we demonstrate repeatable and sharp magnetic field history dependent resistive and superconducting states at zero field. This is markedly different from the expectation of a single non-superconducting state arising from Fig. 1b. The blue arrows in Fig 1b shows the field cycling sequence and amplitudes that are followed to achieve the zero field non-volatile bistable configurations. The magnetic field variations versus time are depicted in the bottom graph of FIG. 1c.

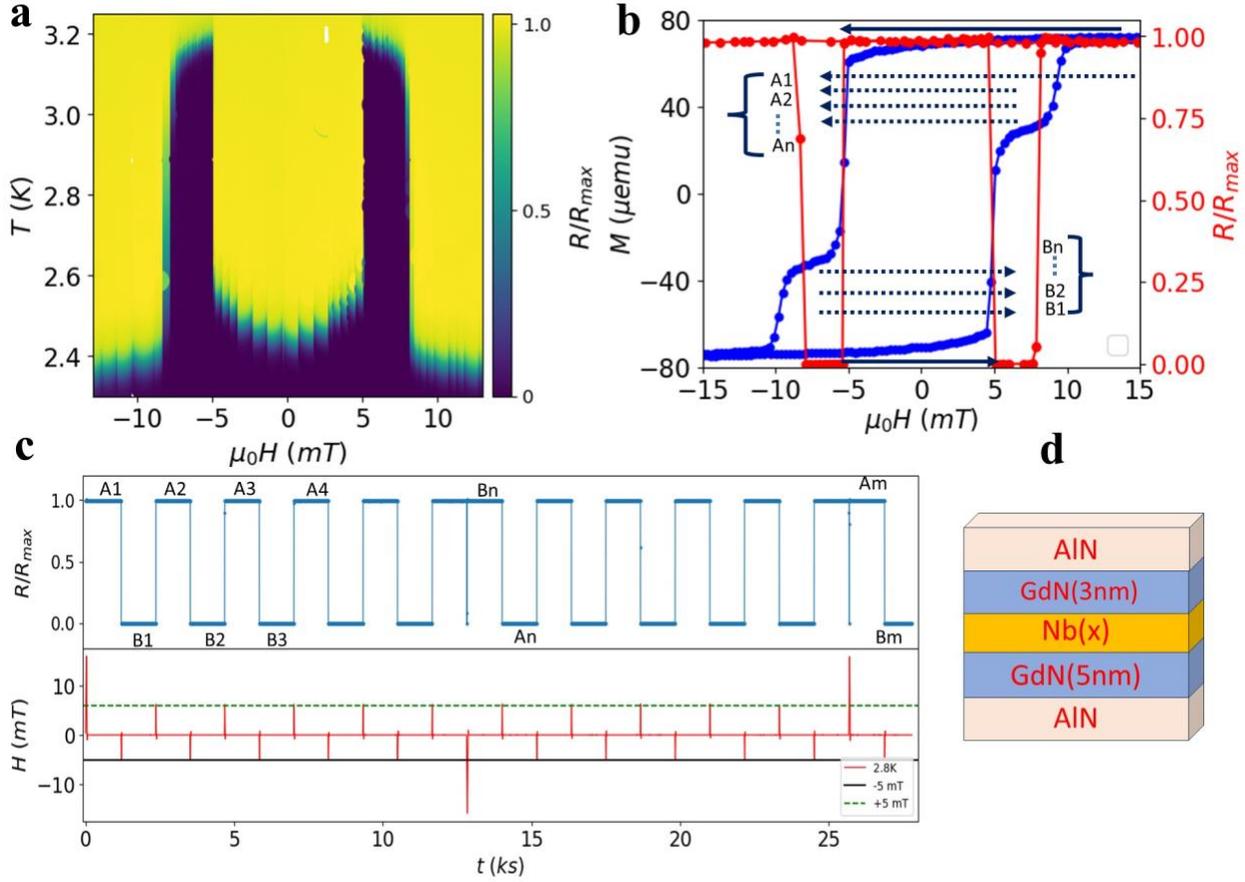

Fig.1| **SEC and bistable configurations in a SSV.** (a) A RT colourmap of SV1 - GdN(5 nm)/Nb(8.5 nm)/GdN(3 nm) in the field range of 0 to +15 mT and 0 to -15mT. (b) The Extracted RH plot from the colormap of SV1 at 2.8 K, along with the magnetic isotherm (M vs H) at 2.8K. (c) Switching of $R/R_{max}$ between "0" and "1" states with symmetric external magnetic fields ($\pm$5mT) shown by the black solid arrow below. The dashed arrows represent asymmetric field cycling routines which are followed for generating figure 3 and 4. (d) Stack structure of all SSVs

Such field history dependent bistable non-volatile states are a direct consequence of SEC. As in plane field is reduced from saturation fields, we enter the anti-parallel (AP) state at $H_{c1} = -5$mT, which triggers superconductivity and the SEC interaction between the top and bottom ferromagnets, and this in turn further stabilises the AP state configuration[10]. From this point onwards, the superconducting state can only be turned off by reversing to the P state configuration by either decreasing field beyond coercive fields of 3nm GdN, or reversing the 5nm GdN with its positive coercive field. As a result, if magnetic field is removed just after entering the AP state, the superconducting state persists even in zero field and in the range $\pm H_{c1}$.

Table 1: List of parameters in SSVs and bilayers for evaluation of CFM state

| Sample Name | Sample Description | $T_C(K)$ Bare Nb | $T(\tau)$ | $\tau(T)$ | $\lambda$ | a(2nm) | a(4nm) |
|---|---|---|---|---|---|---|---|
| SV1* | GdN(3)/Nb(8.5)/GdN(5) | 6.9 | 2.775 | 0.59 | 5.65e$^{-3}$ | 1.369 | 1.82 |
| SV1** | GdN(3)/Nb(8.5)/GdN(5) | 6.9 | 2.775 | 0.59 | 5.65e$^{-3}$ | 1.207 | 2.41 |
| SV2* | GdN(3)/Nb(9)/GdN(5) | 7.1 | 3.438 | 0.51 | 5.55e$^{-3}$ | 1.340 | 2.68 |
| SV2** | GdN(3)/Nb(9)/GdN(5) | 7.1 | 3.438 | 0.51 | 5.55e$^{-3}$ | 1.258 | 2.51 |
| SV3** | GdN(3)/Nb(10)/GdN(5) | 7.2 | 4.05 | 0.43 | 5.53e$^{-3}$ | 1.257 | 2.51 |
| SV4** | GdN(3)/Nb(11)/GdN(5) | 7.5 | 4.71 | 0.37 | 5.42e$^{-3}$ | 1.254 | 2.51 |
| BL** | GdN(3)/Nb(9) | 7.1 | 4.82 | 0.32 | 5.55e$^{-3}$ | 1.115 | NA |

\* Values obtained from the $1^{st}$ order phase transition curve in the 'Maki-Fulde' graph[15].
\*\* Values obtained from the $2^{nd}$ order phase transition curve of the same graph.
N.B. - $T(\tau) = T_{min} + \frac{T_{min} + T_{max}}{2}$
For SV1 and SV2 the values of $\frac{T(\tau)}{T_{c0}}$ is such that both $1^{st}$ and $2^{nd}$ order phase transition of 'Nb' is possible

Since the Tc of these SEC SSVs are greatly reduced, due to presence of large exchange fields; they offer an unique opportunity to test novel phenomena that are predicted to occur at S/F interfaces. Of present interest is a superconductor influenced magnetic state – or the crypto-ferromagnetic (CFM) state which was first discussed by P. W. Anderson, H. Suhl[8] in a magnetic superconductor, and was later generalised to S/F multilayers by Buzdin and Bulaevski[14] and Bergeret et. al[15]. In the CFM state, due to introduction of superconducting condensation energy, domain structures with a length scale smaller than the superconducting coherence length ξ forms in the ferromagnetic layer, and the exchange field will be effectively cancelled over the dimension of the Cooper-pairs, which will reduce the exchange field's detrimental effect on superconductivity.

As per the predictions of Bergeret et al., the possibility of a CFM state is facilitated by three parameters:

$$\lambda = \frac{Jd}{\nu\sqrt{2T_C D^3}} \frac{7\zeta(3)}{2\pi^2}$$

$$a^2 \equiv \frac{2h^2 d^2}{DT_c \eta^2}$$

$$\tau = \frac{T_c - T_\tau}{T_c}$$

Here $J, d, T_C$, and $D$ are exchange stiffness of the ferromagnet, thickness of the ferromagnet, superconducting transition temperature and the diffusion constant of the superconductor respectively. In the second equation, h represents the effective exchange field on the superconductor and is determined using the $T_c$ and $T_\tau$ values and the Maki-Fulde phase diagram of superconductors in an uniform magnetic field[16,17]. $\eta = \frac{v_0^m}{v_0^s} = \sqrt{\frac{E_f^m}{E_f^s}}$, where $v_0^m, v_0^f, E_f^m$ and $E_f^s$ are fermi velocities and fermi energy of the ferromagnet and the superconductor respectively. $T_c$ represents the transition temperature of bare Nb of same thickness, and $T_\tau$ is the temperature where $\tau$ is calculated for that specific sample.

It is evident from the phase diagrams presented by Bergeret et al., that a combination of high $\tau$, low $\lambda$ and intermediate $a$ (0.8 – 1.5) are most favourable for the formation of the CFM state. From Table 1 and comparison with CFM phase diagrams[15], it becomes clear that SV1, SV2, and SV3 are positioned firmly within the CFM region of the phase diagrams with CFM state in the 3nm GdN (effective thickness of 2nm with dead layer of 1nm, as derived from magnetisation switching ratios of 1:2 in Fig. 1b). However, SV4 is possibly located very close to the CFM phase boundary. From this analysis it appears that owing to low $\tau$ (less reduction of Tc due to comparatively lower exchange fields in a bilayer as opposed a tri-layer), the CFM state may not be attainable in a Nb/GdN bilayer sample[18].

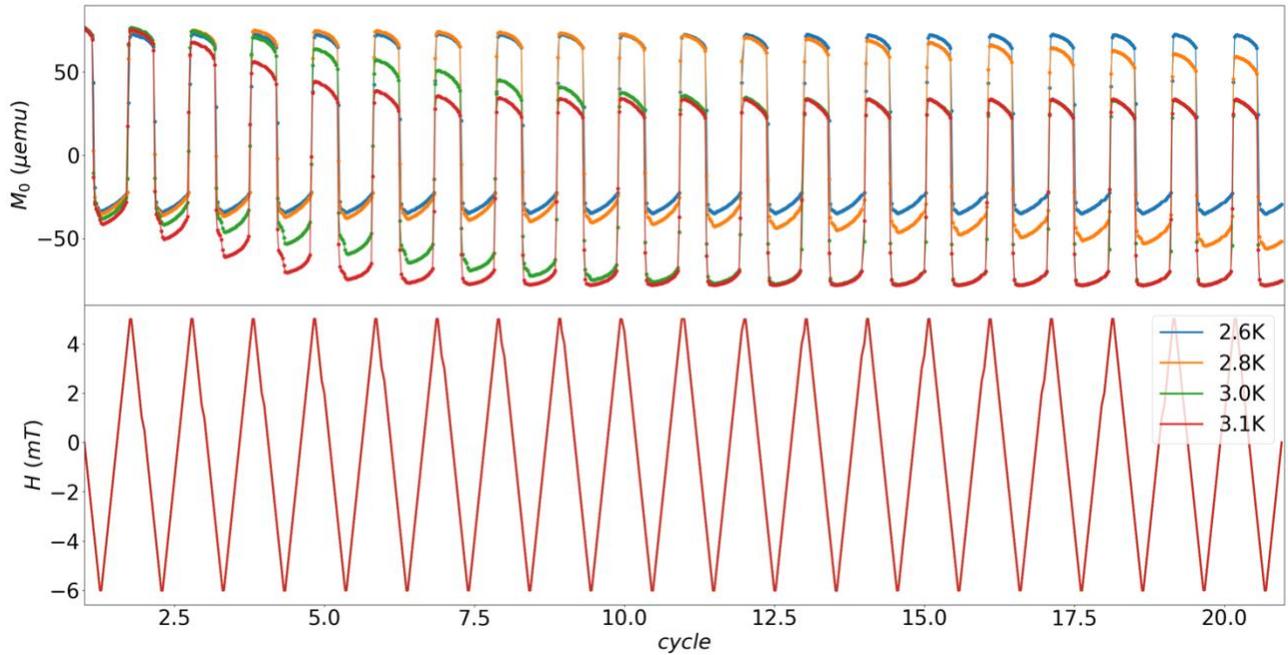

Fig.2:| **Evolution of remnant magnetization as a function of field cycling and temperature.** Evolution of total positive and negative remnant magnetization of SV1 as a function of number of field cycles at 2.6K, 2.8K, 3K and 3.1K. Bottom part shows the amplitude of the positive and negative magnetic fields used during cycling.

To test the above finding of a possible superconductor influenced magnetic state, we measure several MH minor loop cycles of SV1 by applying asymmetric field cycles, where we go deeper into the AF state of the SSV (by applying fields < -5mT) in order to deliberately induce a domain structure in the 3nm GdN layer. At all temperatures, we first saturate both magnets in positive fields, and then perform several asymmetric field cycles. We find that the remnant magnetisation state follows a monotonic temperature dependent magnetisation reduction trajectory. The decrement in magnetisation with cycling is lowest for the lowest measurement temperatures, signifying that a relatively strong superconductor (superconductor with enhanced superconducting condensation energy) tries to dictate the micromagnetic state of the 3nm GdN. This observation is therefore tantamount to direct evidence of a superconductor controlling the micromagnetic state of the ferromagnets.

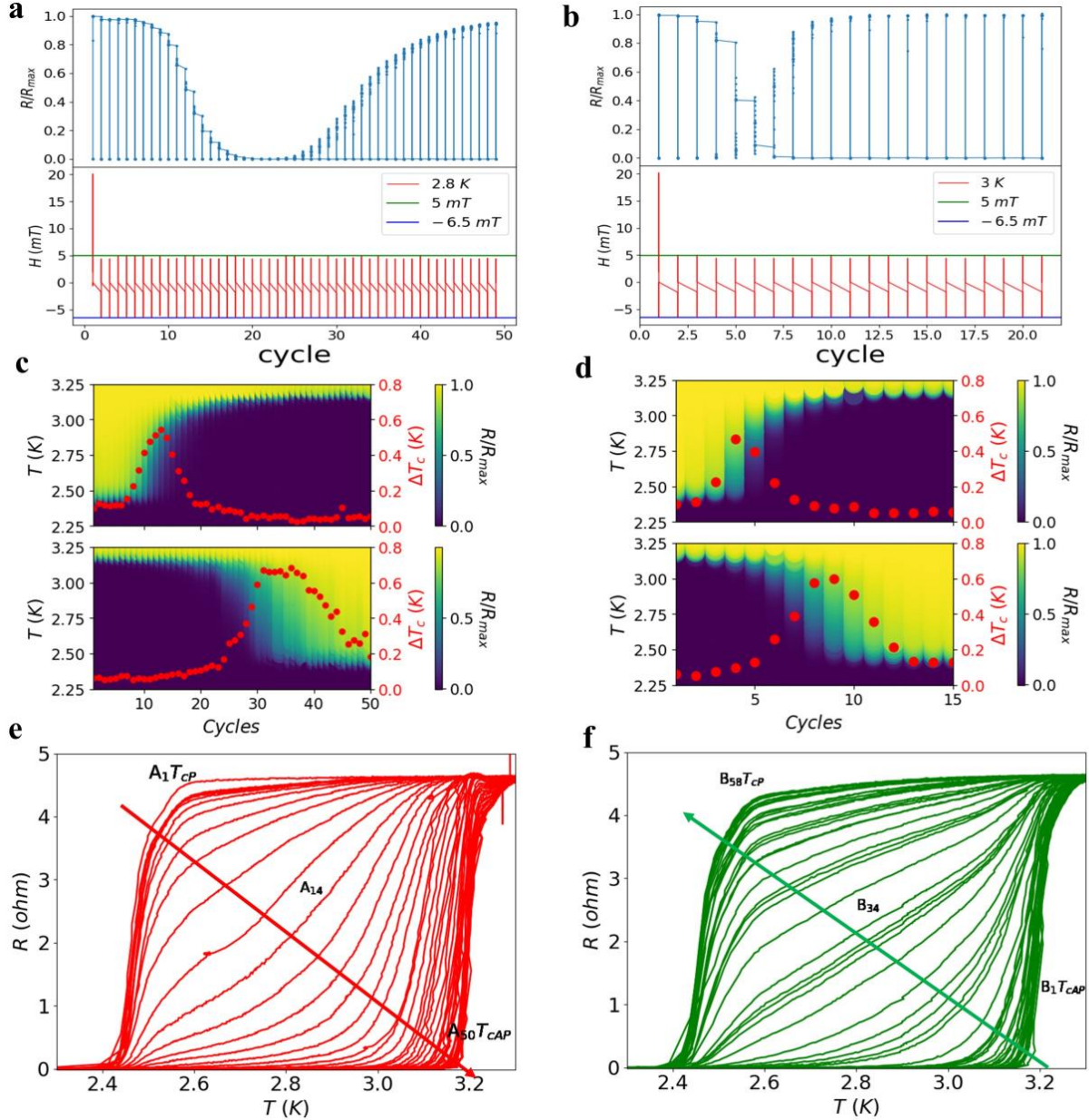

Fig.3| **Reciprocal effect of micromagnetic states on superconductivity.** (a) and (b) represents the normalized resistance of SV1 against number of magnetic cycles at 2.8 K and 3 K. A1, A2, .., An and B1, B2, … , Bn are the zero field states after positive and negative fields respectively after n number of cycles. The red and green curved arrows show the transition of An state to the superconducting state and the transition of the Bn state to the normal state, respectively. (c) Top left – RT colormap of An states at 2.8K as a function of field cycles. Bottom left – RT colormap evolution of Bn states at 2.8K as a function of field cycles. Right hand side Y axis represents the broadening of superconducting transition temperature ($\Delta T_c$) for each An and Bn state. (d) Identical measurements and representation of An and Bn states at 3K. (e) and (f) are the linear representation of An and Bn states in 3c. The red and green arrows represent the direction of evolution of RT curves for A and B states.

Since the evolution of micromagnetic states is found to be a function of temperature dependent asymmetric in-plane magnetic field cycling; we expect a reciprocal action of micromagnetic states on the superconducting state. We therefore perform similar asymmetric field cycling experiments of the SSVs and record the resistivity data to check any such reciprocal action. The result of such asymmetric cycling at 2.8K and 3K for the SV1 spin valve has been illustrated in FIG. 3a and FIG. 3b, respectively. Several zero field non-volatile states in between the superconducting and the normal states are observed, and these zero field states are named as A1, A2, A3,....., An after first, second, third, nth field cycling, respectively after zeroing fields from positive fields. Similarly, B1, B2, B3,.., and Bn are the zero field states after first, second, third, nth field cycling, respectively after zeroing fields from negative side.

The colour maps shown in figure 3c and 3d for a particular temperature are obtained as follows. For Figure 3c, we first stabilise the temperature to 2.8K, then saturate both ferromagnets with positive field and then zero the field. This corresponds to the A1 state at 2.8K. We then measure the Tc of this A1 state through RT measurement, by lowering the temperature to 2.25K and then recording the ultra-slow heating data (heating rate - 0.05K/minute) from 2.25K to 3.25K. After the RT measurement, the temperature is again brought down and stabilised to 2.8K, and a negative field (-6.5mT) is applied, following which the field is zeroed again. This zero field state after negative field is named as the B1 state. The RT plot for B1 is again obtained in an identical manner as A1. Subsequent An states from A2 onwards are the zero field states after application of positive field (+5mT as opposed to positive saturation field for A1 state). We continue such measurements for several cycles to study the nature of A and B states. The major observations from such field cycling experiments and their implications are summarised below:

a) The trajectory of An and Bn state evolution is found to be temperature dependent. This is due to the reciprocal action of micromagnetic states on the superconducting state of the Nb layer in the SSV. This is found in spin valves SV1 to SV4, but is not found in a 3nm GdN/ 9nm Nb sample. This therefore constitutes a control experiment (shown in the Supplementary Information, FIG.3(a)) and provides further evidence in favour of a CFM state in the SSVs.

b) The An states at all temperatures start with the parallel state Tc ($T_{cP}$) and gradually evolve to attain anti-parallel state Tc ($T_{cAP}$). Similarly, Bn states at all temperatures start with the $T_{cAP}$ and gradually evolve to attain $T_{cP}$. However, the number of cycles (C) required for the transition at any particular temperature is different for the A and B states: $C_{B\ states}^{T_{CAP} \to T_{CP}} > C_{A\ states}^{T_{CP} \to T_{CAP}}$. For example, the top graph of FIG. 3c shows that transition from $T_{cP}$ to $T_{cAP}$ only takes a few field cycles (20), but the bottom graph shows that it takes a much larger number of cycles (50) to switch from $T_{cAP}$ to $T_{cP}$. For identical measurements at 3 K, $C_A^{T_{CP} \to T_{CAP}} \cong 5$, whereas $C_B^{T_{CAP} \to T_{CP}} \cong 8$. In other words, if the system starts from the AP state, it prefers to remain longer in AP state. If the system starts from P state, it prefers to transition to the AP state with lesser number of field cyclings. This shows a clear preference of the system for the AP state, which is expected from a SSV system which mediates SEC between the top and bottom ferromagnets[9,10].

c) The stiffness of the AP state is also reflected in the cycle dependent broadness of RT transition (**δ**Tc = $R_N$*0.9 - $R_N$*0.1), presented as filled red circles in Figures 2c and 2d. The corresponding RT plots of all An and Bn states for 2.8K are shown in Figures 3e and 3f. Larger **δ**Tc is observed for greater number of cycles for the Bn states as compared to An states, which is symbolic of system trying to cling onto the AP state Tc. The variations in the broadened critical temperature (ΔTc) after each cycle of magnetic field can be attributed to the interaction between a possible single-domain state of the 5nm GdN layer and the micromagnetic arrangement of the 3nm GdN layer. This interaction probably leads to the occurrence of a spatially varying distribution of SEC between top and bottom GdN layers, which translates to a variation in the value of the superconducting energy gap (Δ(x)) within the superconductor[19]. Consequently, this results in the widening of the ΔTc region. Similar studies involving magnetic field cycling were conducted on spin valves with thicker Nb layers, such as SV2 ,SV3 and SV4, as illustrated in the Supplementary Information (SI, FIG.1, FIG.2 and FIG.3). In cases where the SEC is moderately weak, the likelihood of transitioning between the antiparallel (AP) and parallel (P) states is nearly equal for thicker Nb layers. This results in ΔTc displaying a symmetric behaviour in response to the number of magnetic field cycling steps for SSVs having higher Nb thickness.

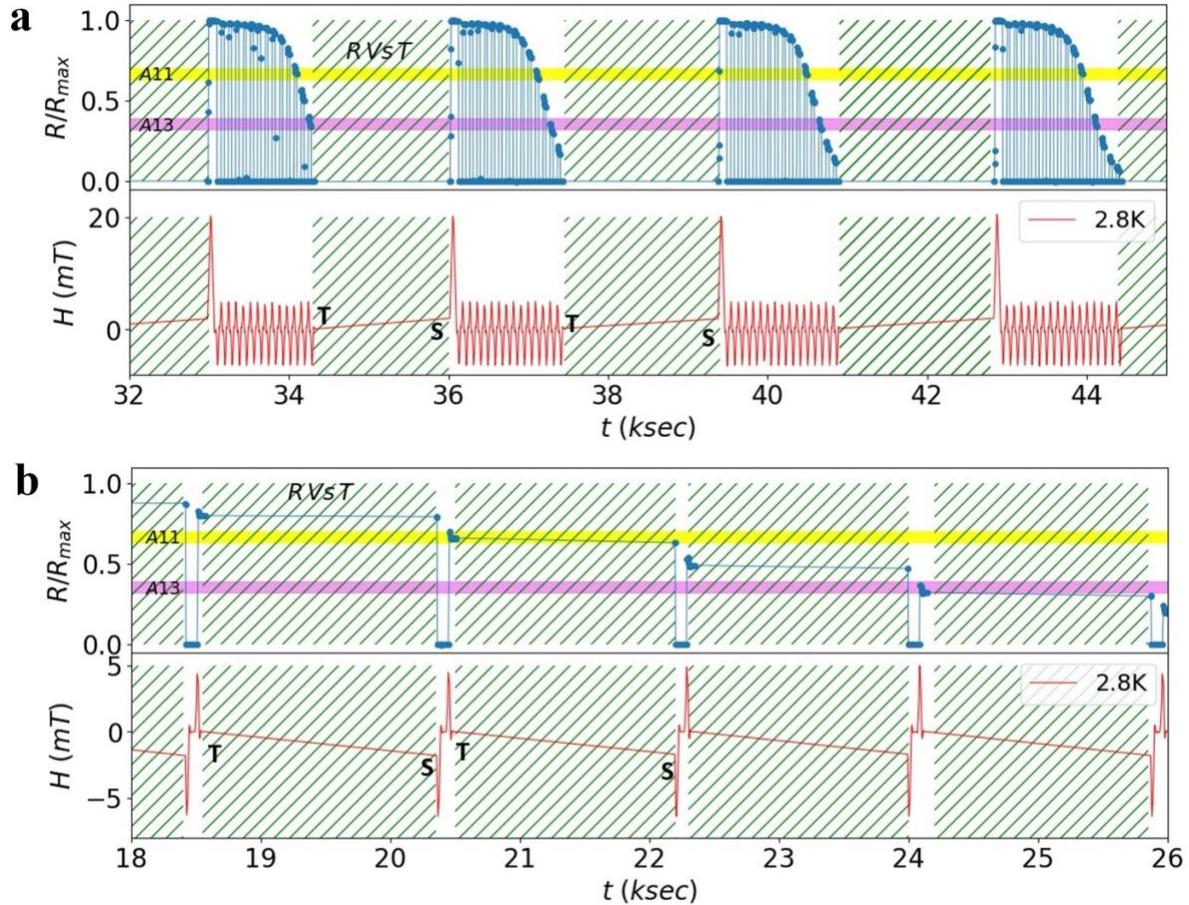

Fig.4| **Reproducibility and non-volatile nature of reciprocally induced resistance states.** (a) Normalized resistance plot of SV1 against number of magnetic cycles (C) at 2.8K followed by RT measurement and repeat of field cycling (C+1 times) starting with positive saturation field every time. A11(Yellow band) and A13(violet band) are the normal states obtained after 11[th] and 13[th] field cycling. The bottom part shows the variation of magnetic field amplitude with time (b) Zoomed view of FIG. 3a around 11[th] and 13[th] magnetic field cycling where RT measurements are performed at zero field after every field cycle. Saturation field has been applied only before generation of A1 state. Bottom part shows the variation of magnetic field with time. 'T' and 'S' points are respectively the magnetic field cycling terminating point and the magnetic field cycling starting point respectively. All hatched portions indicate the time during which the R Vs T measurements of An states were performed.

d) The most striking feature of Figures 3a and 3b is the memory of the device of its preceding An-1 and Bn-1 states, despite the cooling and heating operations performed for the RT measurement between the An-1 (Bn-1) and An (Bn) state measurements. We demonstrate this state memory in Figure 4, by recreating the An and Bn states at 2.8K, using two different measurement protocols. For Figure 4b, we use the same protocol as used for Figures 3a,b. However, for Figure 4a, we create the states afresh each time after every RT measurement. This entails starting with positive field saturation followed by n field cycling between -6.5mT and +5mT and then performing an RT measurement every single time. Identical RT colormaps (corresponding to Figure 3c for 2.8K) are obtained in either case, with Tc and Resistance of every An and

Bn state remaining exactly identical in both kind of measurement protocols. As an illustrative example, we mark the resistance of A11 and A13 states in both measurement protocols in Figure 4a and 4b (The reproducibility of A11 and A13 states for larger number of cycles or larger time is shown in the SI, FIG. 5).

We find that this above-described state memory exists for all temperatures between $T_{CP}$ and $T_{CAP}$. Such a device which demonstrates sharp switching from the 0 to 1 state and vice versa (Fig. 1) and can also give rise to reproducible, non-volatile and temporally non decaying multiple resistance states between 0 and 1 states make these SSVs great candidates for in-memory cryogenic computing.

In conclusion, we have demonstrated that in SSVs with high reduced transition temperatures, owing to the possible formation of a crypto-ferromagnetic state, the strength of the superconducting state can dictate the nature of domain state in a thin ferromagnetic layer. This leads to a reciprocal effect of the magnetic state on the superconducting state, which can be evinced through asymmetric field cycling and resistivity measurements of the SSV. Harnessing the power of both symmetric and asymmetric field cycling, we achieve either reproducible, time invariant and non-volatile bistable states; or reproducible, time invariant, non-volatile, multiple resistance states respectively. Such a device demonstration paves the way towards a novel approach to cryogenic neuromorphic computing.

**Methods**
All layers of the SSV were deposited using an UHV sputtering system, with base pressure of $10^{-9}$ hPa, SSV trilayers were grown on n-doped Si substrates with a 285nm thick $SiO_2$ thermally grown oxide. The bottom GdN layer was grown on top of a buffer layer of AlN (10nm) and the trilayer was capped with another AlN layer (20nm) at the top. All layers were deposited without breaking the vacuum. All 4 SSVs (SSV1-4) were deposited in the same sputtering run, by means of a substrate carrying rotating table arrangement and Nb thickness was varied using different rotation speeds for different samples. Rotation of the substrate table was achieved using a computer controlled stepper motor. Magnetisation measurements as a function of in plane magnetic fields were performed on a AlN/GdN(3nm)/AlN sample using a M/s Quantum Design SQUID magnetometer system. For the magneto-transport measurements, the SSVs were wire bonded in a 4-point geometry, to a custom-made PCB devoid of any magnetic material, which were then affixed to a probe that was lowered into a variable temperature insert of an Oxford Teslatron Pulse tube Cryostat. In plane magnetic fields were applied using the inbuilt superconducting solenoid of the Teslatron system. The QCODES platform was adopted for data acquisition.

**Acknowledgements**
We express our sincere thanks to Prof. P Raychaudhuri for helping us carry out the SQUID VSM measurements at TIFR, Mumbai. The work was financially supported by a Core Research Grant from Department of Science and Technology, India, file number CRG/2019/004758.

**Author contributions**
BD, SB and AP conceived the SSV structures. SB grew the SSV multilayers, and performed measurements for all SSVs corresponding to Fig 1a. BD and AP conceived all experiments related to field cycling and investigation of CFM state. PB helped in performing the field cycling experiments. BD and AP wrote the manuscript. All authors discussed the results and commented on the manuscript.

**References:**

1. Buzdin, A., Bulaevskii, L. & Panyukov, S. Critical-current oscillations as a function of the exchange field and thickness of the ferromagnetic metal (F) in an SFS Josephson junction. *JETP Lett* 178–180 (1982).

2. Bergeret, F. S., Volkov, A. F. & Efetov, K. B. Long-Range Proximity Effects in Superconductor-Ferromagnet Structures. *Phys. Rev. Lett.* **86**, 4096–4099 (2001).

3. Bergeret, F. S., Volkov, A. F. & Efetov, K. B. Odd triplet superconductivity and related phenomena in superconductor-ferromagnet structures. *Rev. Mod. Phys.* **77**, 1321–1373 (2005).

4. Ryazanov, V. V. *et al.* Coupling of Two Superconductors through a Ferromagnet: Evidence for a $\pi$ Junction. *Phys. Rev. Lett.* **86**, 2427–2430 (2001).

5. Kontos, T. *et al.* Josephson Junction through a Thin Ferromagnetic Layer: Negative Coupling. *Phys. Rev. Lett.* **89**, 137007 (2002).

6. Robinson, J. W. a, Witt, J. D. S. & Blamire, M. G. Controlled injection of spin-triplet supercurrents into a strong ferromagnet. *Science* **329**, 59–61 (2010).

7. Khaire, T. S., Khasawneh, M. A., Pratt, W. P. & Birge, N. O. Observation of Spin-Triplet Superconductivity in Co-Based Josephson Junctions. *Phys. Rev. Lett.* **104**, 137002 (2010).

8. Anderson, P. W. & Suhl, H. Spin Alignment in the Superconducting State. *Phys. Rev.* **116**, 898–900 (1959).

9. De Gennes, P. G. Coupling between ferromagnets through a superconducting layer. *Phys. Lett.* **23**, 10–11 (1966).

10. Zhu, Y., Pal, A., Blamire, M. G. & Barber, Z. H. Superconducting exchange coupling between ferromagnets. *Nat. Mater.* **16**, 195–199 (2017).

11. Sebastian, A., Le Gallo, M., Khaddam-Aljameh, R. & Eleftheriou, E. Memory devices and applications for in-memory computing. *Nat. Nanotechnol.* **15**, 529–544 (2020).

12. Tagirov, L. R. Low-Field Superconducting Spin Switch Based on a Superconductor / Ferromagnet Multilayer. *Phys. Rev. Lett.* **83**, 2058–2061 (1999).

13. Kushnir, V. N., Sidorenko, A., Tagirov, L. R. & Kupriyanov, M. Y. Basic superconducting spin valves. *Nanosci. Technol.* 1–29 (2018). doi:10.1007/978-3-319-90481-8_1

14. Buzdin, A. & Bulaevskii, L. N. Ferromagnetic film on the surface of a superconductor: possible onset of inhomogeneous magnetic ordering. *J. Exp. Theor. Phys.* **67**, 576–578 (1988).

15. Bergeret, F. S., Efetov, K. B. & Larkin, A. I. Nonhomogeneous magnetic order in superconductor-ferromagnet multilayers. *Phys. Rev. B* **62**, 11872–11878 (2000).

16. Maki, K. The Behavior of Superconducting Thin Films in the Presence of Magnetic Fields and Currents. *Progress of Theoretical Physics* **31**, 731–741 (1964).

17. Fulde, P. & Ferrell, R. A. Superconductivity in a Strong Spin-Exchange Field. *Phys. Rev.* **135**, A550–A563 (1964).

18. Banerjee, P., Sharma, P. K., Bhakat, S., Dutta, B. & Pal, A. Large tuneable exchange fields due to purely paramagnetically limited domain wall superconductivity. (2023).

19. Benfatto, L., Castellani, C. & Giamarchi, T. Broadening of the Berezinskii-Kosterlitz-Thouless superconducting transition by inhomogeneity and finite-size effects. *Phys. Rev. B* **80**, 214506 (2009).

# Supplementary Information

## Bistable switching and multiple states in SV2

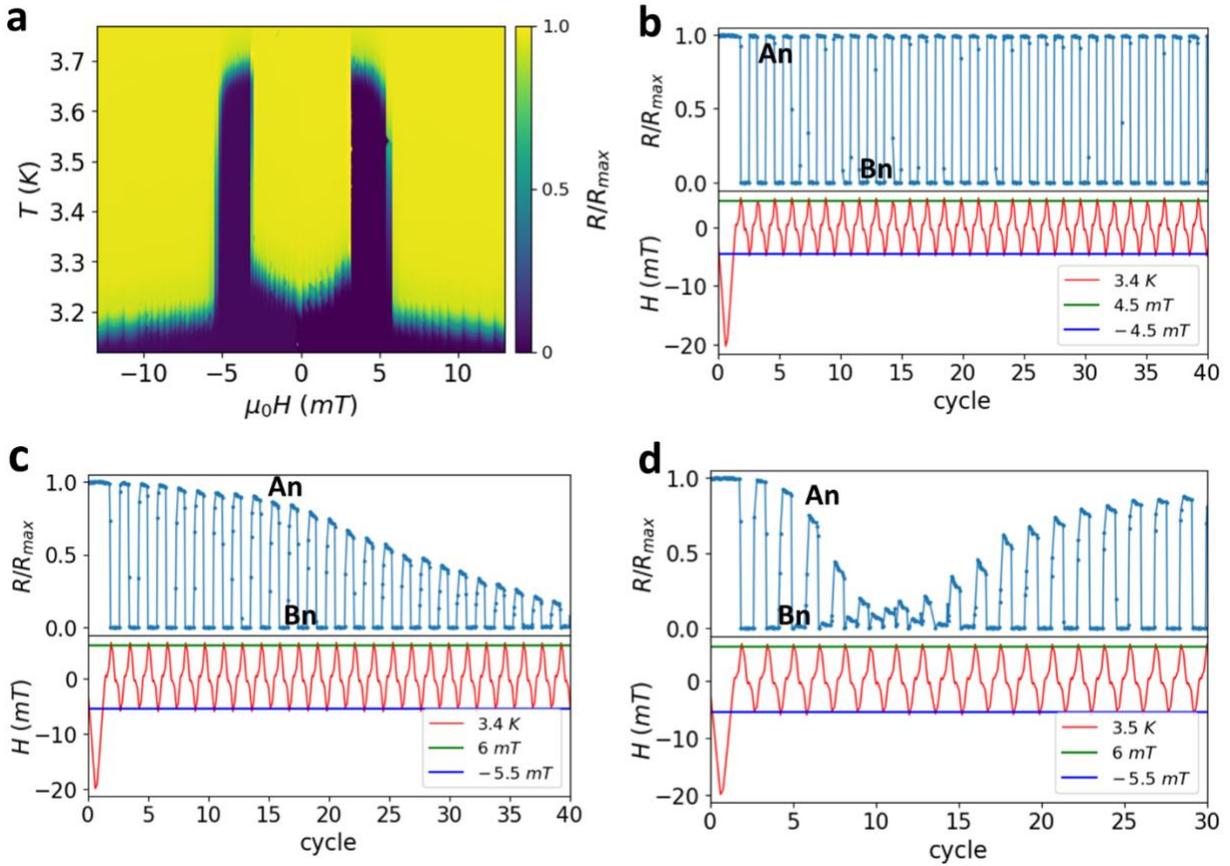

Figure S1:(a) The evolution of the RT color map of the SV2 at different magnetic field. (b) Switching of $R/R_{max}$ between "0" and "1" states with symmetric external magnetic fields ($\pm$ 4.5 mT). (c) and (d) show the trajectory of the A and B states of SV2 as a function of the number of asymmetric magnetic field cycling at temperatures 3.4 K and 3.5 K, respectively. The states labeled as An and Bn correspond to the parallel and antiparallel state after undergoing 'n' number of magnetic field cycling.

# Bistable switching and multiple states in SV3

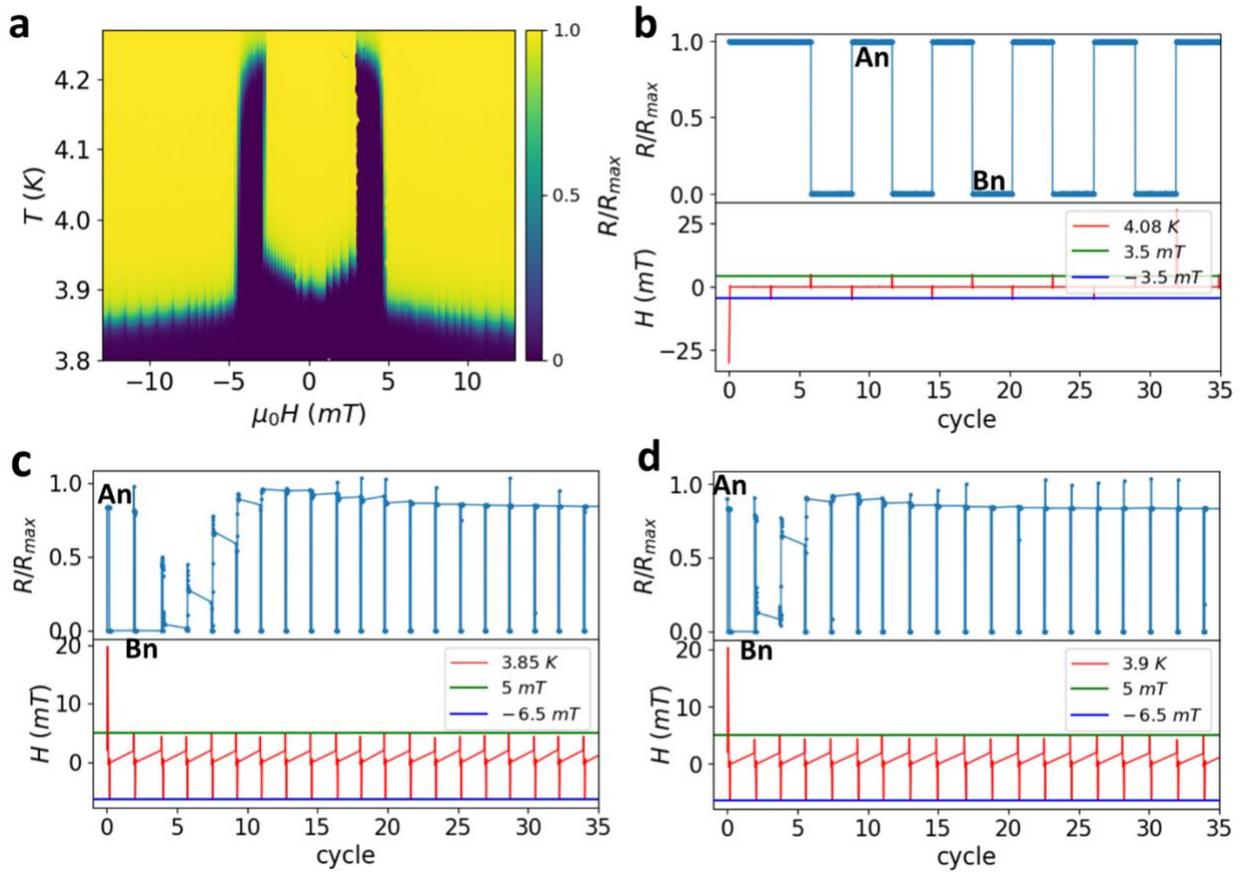

Figure S2: (a) The evolution of the RT color map of the SV3 at different magnetic field. (b) Switching of $R/R_{max}$ between "0" and "1" states with symmetric external magnetic fields ($\pm$ 3.5 mT). (c) and (d) show the trajectory of the A and B states of SV3 as a function of the number of asymmetric magnetic field cycles at temperatures 3.85 K and 3.9 K, respectively. The states labeled as An and Bn correspond to the parallel and antiparallel state after undergoing 'n' number of magnetic field cycling.

## Bistable switching and multiple states in SV4

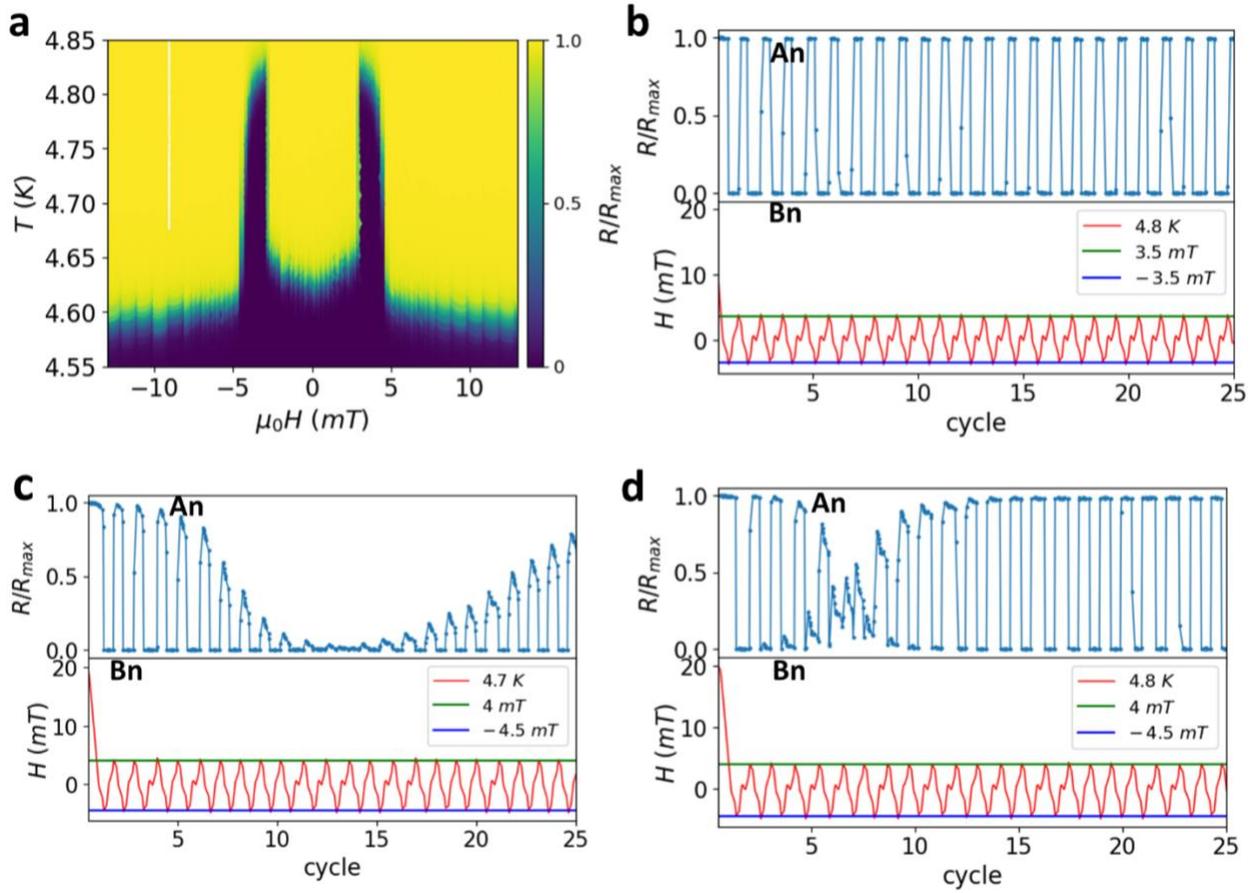

Figure S3: (a) The evolution of the RT color map of the SV4 at different magnetic field. (b) Switching of $R/R_{max}$ between "0" and "1" states with symmetric external magnetic fields ($\pm$ 3.5 mT). (c) and (d) show the trajectory of the A and B states of SV4 as a function of the number of asymmetric magnetic field cycles at temperatures 4.7 K and 4.8 K, respectively. The states labeled as An and Bn correspond to the parallel and antiparallel state after undergoing 'n' number of magnetic field cycling.

**Minor loop cycling of a GdN(3nm)/Nb(5nm) sample**

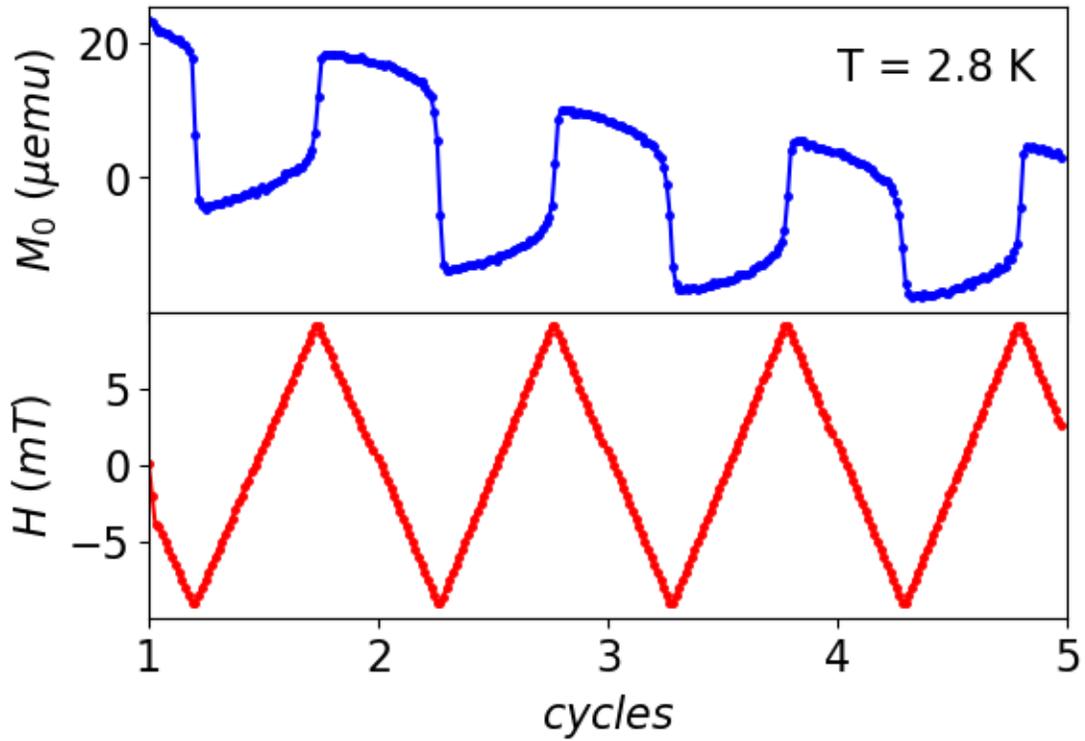

Figure S4: A magnetic field cycling measurement of GdN(3nm)/Nb(5nm) bilayer sample between $\pm H_{c1}$, which indicates the decay of the magnetization value of 3nm GdN layer with magnetic field cycling in the absence of CFM state and the SEC.

**Complete measurement for excerpt taken in Figure 4a of main manuscript**

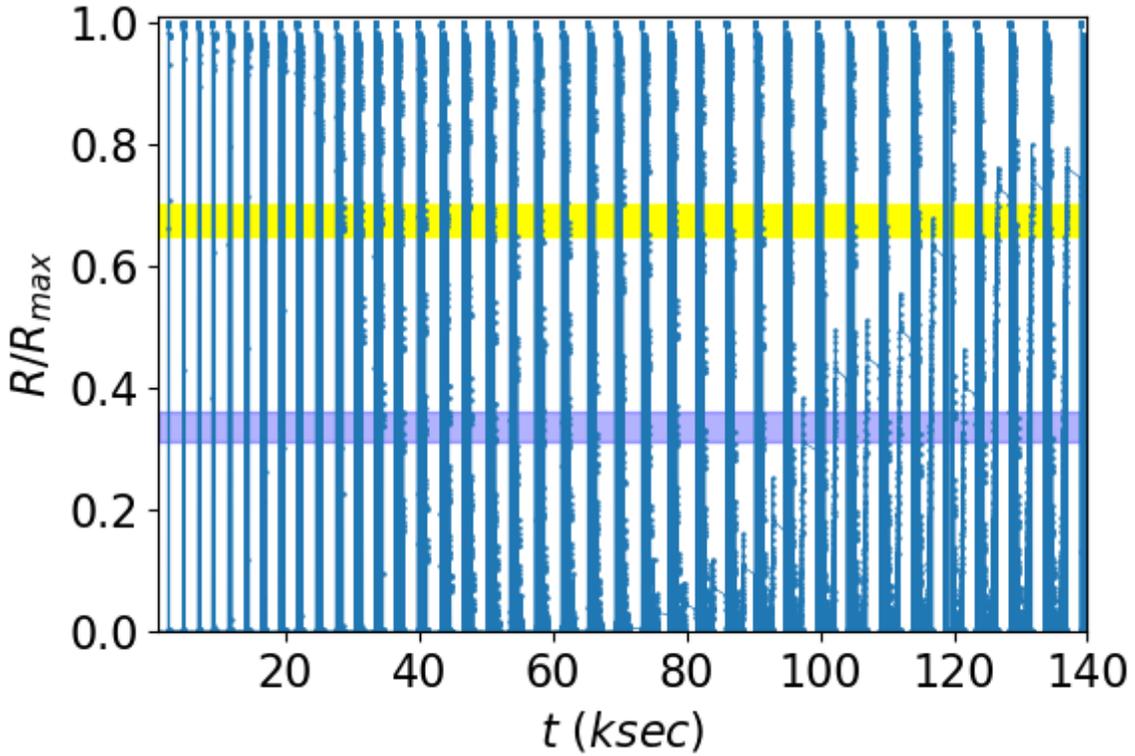

Figure S5: The recurring pattern of An (or Bn) states can be observed after subjecting them to 'n' number of cycles of magnetic fields. To illustrate, the progression of A11 (Yellow band) and A13 (blue band) states following the 11th and 13th rounds of magnetic field cycling has been demonstrated. The width represents the margin of error, primarily influenced by the precision of both magnetic field and resistance measurements.